\documentstyle[12pt]{article}
\begin{document}
\baselineskip=24pt
\def\rd{{\rm d}}
\newcommand{\Lamb}{\Lambda_{b}}
\newcommand{\Lamc}{\Lambda_{c}}
\newcommand{\dsp}{\displaystyle}
\newcommand{\nn}{\nonumber}
\newcommand{\dfr}[2]{ \displaystyle\frac{#1}{#2} }
\newcommand{\Lag}{\Lambda \scriptscriptstyle _{ \rm GR} } 
\newcommand{\pa}{p\parallel}
\newcommand{\pe}{p\perp}
\newcommand{\pet}{p\top}
\newcommand{\paa}{p'\parallel}
\newcommand{\pee}{p'\perp} 
\newcommand{\pete}{p'\top} 
\renewcommand{\baselinestretch}{1.5}
\begin{titlepage}
\vspace{-20ex}
\begin{flushright}
\vspace{-3.0ex} 
    \it{HUPD-9610} \\
\vspace{-2.0mm}    
       \it{April, 1996}\\
\vspace{-2.0mm}
\vspace{5.0ex}
\end{flushright}

\centerline{\Large\sf Isgur-Wise Function for $\Lamb \rightarrow
\Lamc$ in B-S Approach}
\vspace{6.4ex}
\centerline{\large\sf  	X.-H. Guo$^{1,2}$ and T. Muta$^{1}$}
\vspace{3.5ex}
\centerline{\sf 1. Department of Physics, Hiroshima University,  Japan}
\centerline{\sf 2. Institute of High Energy Physics, Academia Sinica,
Beijing, China}
\vspace{6ex}
\begin{center}
\begin{minipage}{5in}
\centerline{\large\sf 	Abstract}
\vspace{1.5ex}
\small {In the heavy quark limit, the heavy baryon $\Lambda_Q$ (Q=b or
c) can be regarded as composed of a heavy quark and 
a scalar light diquark which has good spin and flavor quantum numbers.
Based on this picture we establish the Bethe-Salpeter (B-S) 
equation for $\Lambda_Q$ in the leading order of $1/m_Q$ expansion.
With the kernel containing both the scalar confinement and
one-gluon-exchange terms we solve the B-S equation numerically.
The Isgur-Wise function for  $\Lamb \rightarrow \Lamc$ is obtained
numerically from our model. Comparison with other model calculations
are also presented. It seems that the Isgur-Wise function for $\Lamb
\rightarrow \Lamc$  drops faster than that for $B \rightarrow D$. The
differential and total decay widths for  $\Lamb \rightarrow \Lamc l
\bar{\nu}$ are given in the limit $m_{b,c}\rightarrow \infty$.}

\end{minipage}
\end{center}

\vspace{1cm}

\end{titlepage}
\vspace{0.2in}
{\large\bf I. Introduction}
\vspace{0.2in}

The past few years have seen much progress in heavy flavor physics due 
to the discovery of the new flavor and spin symmetries $SU(2)_f \times 
SU(2)_s$ in the heavy quark limit and the establishment of the heavy quark 
effective theory (HQET) \cite{wise}. In the framework of the HQET  
the processes involving heavy quarks can be simplified since the HQET may 
reduce the number of form factors. For instance in the
leading order of $1/m_Q$ expansion only one form factor (the
Isgur-Wise function) remains for $\Lamb \rightarrow \Lamc$
transition. However, to get the complete physics one still has to face 
the basic problem, the nonperturbative effects of QCD.
As the formally exact equation to describe the relativistic bound
state, the B-S equation \cite{bs}\cite{lurie} can be simplified to a 
great extent in the
heavy quark limit and has been applied to give many theoretical
results concerning heavy mesons \cite{dai}. At present, there are not
many experimental results for heavy baryons. But in the future we may
expect more and more data coming out in LEP and
the forthcoming B-factory. Actually recently OPAL has measured some physical
quantities for $\Lamb$ such as the lifetime of $\Lamb$ and the product 
branching ratio for $\Lamb \rightarrow \Lambda l^- \bar{\nu} X$
\cite{opal}. Hence the study of heavy baryons are of particular
interestes in the near future. It is the motivation of
the present paper to establish the B-S equation for the heavy baryon
in the heavy quark limit and then solve it numerically by assuming
some reasonable form of its kernel, and consequently give some
phenomenological predictions.

When the quark mass is very heavy comparing with the QCD scale
$\Lambda_{QCD}$, the light degrees of freedom in a heavy baryon  
$\Lambda_{Q}$ (Q=b or c) becomes blind to the flavor and spin quantum
numbers of the heavy quark because of the $SU(2)_f \times 
SU(2)_s$ symmetries. Therefore, the angular momentum  
and flavor quantum numbers 
of the light degrees  of freedom (the light diquark) become good quantum 
numbers which can be used to classify heavy baryons. For example, 
$\Lambda_Q$ and $\Sigma_{Q}^{(*)}$ correspond to the zero and one
angular momentum states of the light diquark respectively. Hence it is natural 
to regard the heavy baryon to be composed of a heavy quark and a light 
diquark. In the present paper we concentrate on $\Lambda_{Q}$ which has
a scalar light diquark $S_{[ud]}$ with [ud] flavor quantum number and 
zero spin and isospin. Other heavy baryons like $\Sigma_{Q}^{(*)}$ can 
be dealt with in the similar way. 

Based on the above picture of the composition of the heavy baryon the
three body system is simplified to two body system. We will establish the 
B-S equation for the heavy baryon in this picture. Then we will solve
this equation numerically by assuming that the kernel contains the 
scalar confinement and
one-gluon-exchange terms. Then we will apply the B-S equation to
calculate the Isgur-Wise function for $\Lamb \rightarrow \Lamc$ and
the decay width for the semileptonic decay of $\Lamb$ to $\Lamc$ in
the order $m_{b,c} \rightarrow \infty$.

The remainder of this paper is organized as follows. In Sect. II we
establish the B-S equation for the heavy quark and light scalar
diquark system and discuss the form of its kernel. Consequently this
equation can be solved numerically.
Then in Sect. III we apply the solution of the B-S equation to
calculate the Isgur-Wise function for $\Lamb \rightarrow \Lamc$ and
also the differential and total decay widths for $\Lamb \rightarrow \Lamc l
\bar{\nu}$. Comparison with other model calculations for the
Isgur-Wise function will also be presented. Finally,
Sect. VI is served for summary and discussions.

\vspace{0.2in}
{\large\bf II. The B-S equation for $\Lambda_Q$}
\vspace{0.2in}

As discussed in Introduction $\Lambda_Q$ is regarded as the bound state
of a heavy quark and a light scalar diquark. Based on this picture 
we can define the B-S wave function of $\Lambda_{Q}$ by
\begin{equation}
\chi(x_1, x_2, P)=<0|T \psi(x_1) \varphi(x_2)|\Lambda_Q (P)>,
\label{2a}
\vspace{2mm}
\end{equation}
where $\psi(x_1)$ and $\varphi(x_2)$ are the wave fuctions of the heavy
quark and the light scalar diquark repectively, $P=m_{\Lambda_{Q}}v$
is the momentum of
$\Lambda_Q$ and $v$ is its velocity. Let $m_Q$ and
$m_D$  be the masses of the heavy quark and the light diquark in the baryon, 
$\lambda_1=\frac{m_Q}{m_Q+m_D}, \lambda_2=\frac{m_D}{m_Q+m_D}$, $p$ be the 
relative momentum of the two constituents. The B-S wave function in
the momentum space is defined as
\begin{equation}
\chi(x_1, x_2, P)=e^{iPX}\int \frac{\rd^4 p}{(2\pi)^4}e^{ipx}\chi_P(p),
\label{2b}
\vspace{2mm}
\end{equation}
where $X=\lambda_1 x_1+\lambda_2 x_2$ is the coordinate of the center of 
mass and $x=x_1-x_2$. The momentum of the heavy quark is
$p_1=\lambda_1 P+p$ and that of the diquark is $p_2=-\lambda_2 P+p$.

The derivation of the B-S equation for the two fermion systems can be
found in textbooks \cite{lurie}. In the same way we can prove that for 
the fermion and scalar object 
system the form of the B-S equation is still valid.
$\chi_P(p)$ satisfies the following B-S equation
\begin{equation}
\chi_P(p)=S_F(\lambda_1 P+p)\int \frac{\rd^4q}{(2\pi)^4}G(P,p,q)\chi_P(q)
S_D(-\lambda_2 P+p),
\label{2bb}
\vspace{2mm}
\end{equation}
where $G(P,p,q)$ is the kernel which is defined as the sum of the two 
particle irreducible diagrams. 
In the following we will use the variables $p_l=v\cdot p
-\lambda_2 m_{\Lambda_Q}, p_t=p-(v\cdot p)v$. Then in the leading
order of $1/m_Q$ expansion we have
\begin{eqnarray}
S_F(\lambda_1 P+p)&=&\frac{i(1+\rlap/v)}{2(p_l+E_0+m_D+i\epsilon)}, \nn \\
S_D(-\lambda_2 P+p)&=&\frac{i}{p_{l}^{2}-W_{p}^{2}+E_0+i\epsilon},
\label{2c}
\vspace{2mm}
\end{eqnarray}
where $W_{p}=\sqrt{p_{t}^{2}+m_{D}^{2}}$ and $E_0$ is the binding energy.
The corrections to eq. (\ref{2c}) are from $O(1/m_Q)$.

In general, $\chi_P(p)$ can be expanded as the following form,
\begin{equation}
\chi_P(p)=(A+B\rlap/v +C\rlap/p +D\rlap/v \rlap/p)u_{\Lambda_Q}(v,s),
\label{2d}
\vspace{2mm}
\end{equation}
where $u_{\Lambda_Q}(v,s)$ is the spinor of $\Lambda_Q$ with helicity
s and A, B, C, D are Lorentze scalar functions.

From eqs. (\ref{2bb}) and (\ref{2c}) it can be seen that 
\begin{equation}
\rlap/v \chi_P(p)=\chi_P(p).
\label{2e}
\vspace{2mm}
\end{equation}

Combining eqs. (\ref{2d}) and (\ref{2e}) we immediately see that
\begin{equation}
\chi_P(p)=\phi_P(p)u_{\Lambda_Q}(v,s),
\label{2f}
\vspace{2mm}
\end{equation}
where $\phi_P(p)$ is a scalar field. This form is consistent with our
picture about the heavy baryon. 

We assume the kernel has the  form
\begin{equation}
-iG=I\otimes I V_1 +v_{\mu} \otimes (p_2+p'_2)^{\mu} V_2, 
\label{2g}
\vspace{2mm}
\end{equation}
where the first term arises from scalar confinement and the second one 
is from one gluon exchange diagram. The vertex of gluon with two scalar
diquarks is proportional to $(p_2+p'_2)_\mu$ ($p_2$ and $p'_2$ are the 
momenta of the two scalar diquarks on the vertex) and a form factor 
$F((p_2+p'_2)^2)$ (see Fig. 1). Substituting eqs. (\ref{2c}), (\ref{2f}) and
(\ref{2g}) into (\ref{2bb}) we have
\begin{equation}
\phi_P(p)=-\frac{1}{(p_l+E_0+m_D+i\epsilon)(p_{l}^{2}-W_{p}^{2}+i\epsilon)} 
\int \frac{\rd^4q}{(2\pi)^4}G(P,p,q)\phi_P(q).
\label{2h}
\vspace{2mm}
\end{equation}

Now consider the vertex of two heavy quarks with gluon. The momenta of 
the two heavy quarks are $p_1=\lambda_1 m_{\Lambda_Q}v+p$ and
$p'_1=\lambda_1 m_{\Lambda_Q}v+q$ respectively. p 
and q are of the order $\Lambda_{QCD}$. In the heavy quark 
limit the heavy quark is almost on-shell, therefore, $p_l=q_l$ on this
vertex. Hence we can make the convariant instantaneous approximation
\cite{dai} in the kernel. Then eq. (\ref{2h}) becomes
\begin{equation}
\phi_P(p)=-\frac{i}{(p_l+E_0+m_D+i\epsilon)(p_{l}^{2}-W_{p}^{2}+i\epsilon)} 
\int \frac{\rd^4q}{(2\pi)^4}(\tilde{V}_1 +2p_l \tilde{V}_2)\phi_P(q),
\label{2i}
\vspace{2mm}
\end{equation}
where $\tilde{V}$ is defined as $\tilde{V}\equiv V|_{p_l=q_l}$.

In general $\phi_P(p)$ can be the function of $p_l$ and $p_t$. Defining
$\tilde{\phi}_P(p_t)=\int \frac{\rd p_l}{2\pi} \phi_P(p)$ one gets
immediately the B-S equation for $\tilde{\phi}_P(p_t)$
\begin{equation}
\tilde{\phi}_P(p_t)=-\frac{1}{2(E_0-W_p+m_D)W_{p}} 
\int \frac{\rd^3q_t}{(2\pi)^3}(\tilde{V}_1 -2W_p \tilde{V}_2)\tilde{\phi}_P(q_t).
\label{2j}
\vspace{2mm}
\end{equation}

If one knows the form for the kernel $\tilde{V}_1$ and $\tilde{V}_2$, 
then $\tilde{\phi}_P(p_t)$ can be obtained. Consequently $\phi_P(p)$
can be solved since after integrating $q_l$ in eq. (\ref{2i}) we get
the relation between $\phi_P(p)$ and $\tilde{\phi}_P(q_t)$
\begin{equation}
\phi_P(p)=\frac{i}{(p_l+E_0+m_D+i\epsilon)(p_{l}^{2}-W_{p}^{2}+i\epsilon)} 
\int \frac{\rd^3q_t}{(2\pi)^3}(\tilde{V}_1 +2p_l \tilde{V}_2)
\tilde{\phi}_P(q_t).
\label{2k}
\vspace{2mm}
\end{equation}

The kernel $\tilde{V}_1$ and $\tilde{V}_2$ for the B-S equation in the
meson case was given in \cite{dai} as 
\begin{equation}
\tilde{V}_1\mid _{\rm
meson}=\frac{8\pi\kappa'}{[(p_t-q_t)^2+\mu^2]^2}-(2\pi)^3
\delta^3  (p_t-q_t)
	\int \frac{\rd^3k}{(2\pi)^3}\frac{8\pi\kappa'}{(k^2+\mu^2)^2}, 
\label{2la}
\vspace{2mm}
\end{equation}
\begin{equation}
\tilde{V}_2\mid _{\rm meson}=-\frac{16\pi}{3}
	\frac{\alpha_{s eff}}{(p_t-q_t)^2+\mu^2},
\label{2lb}
\vspace{2mm}
\end{equation}
where $\kappa'$ and $\alpha_{s eff}$ are coupling parameters related
to scalar confinement and one-gluon-exchange diagram
respectively. The second term in eq. (\ref{2la}) is the counter term
which removes the infra-red divergence in the integral equation.
From potential model $\kappa'$ is from 0.18 to 0.2.
The parameter $\mu$ is introduced to avoid the infra-red divergence in 
numerical calculations. The limit $\mu \rightarrow 0$ is taken in the end.

In the baryon case, since confinement is still due to scalar interaction the
form of $\tilde{V}_1$ need not be changed. Only the parameter
$\kappa'$ has to be replaced by $\kappa$ which describes the
confinement interaction between the quark and diquark. However, since
the diquark is not a point-like object, there should be a form factor
$F(Q^2) (Q=p_2-p'_2)$ to describe the vertex of the gluon and two scalar
diquarks. In general, this vertex is (Fig. 1)
$$ ig_s \frac{\lambda^a}{2} (p_2+p'_2)^\mu F(Q^2)$$
with $\lambda$'s being the Gell-Mann color matrices and $g_s$ the
strong coupling constant. The form of
$F(Q^2)$ was assumed as \cite{kroll}   
\begin{equation}
F(Q^2)=\frac{\alpha_{s eff}Q_{0}^{2}}{Q^{2}+Q_{0}^{2}},
\label{2m}
\vspace{2mm}
\end{equation}
where $Q_{0}^{2}$ is a parameter which freezes $F(Q^2)$ when $Q^2$ is
very small. In the high energy region the form factor is proportional
to $1/Q^2$ which is consistent with perturbative QCD calculations
\cite{brodsky}. By analyzing the eletromagnetic form factor
for the proton it was found that $Q_{0}^{2}=3.2$GeV$^2$ can lead to
consistent results with the experimental data \cite{kroll}.

Based on the above analysis the kernel for the B-S equation in the
baryon case is of the following form
\begin{eqnarray}
\tilde{V}_1&=&\frac{8\pi\kappa}{[(p_t-q_t)^2+\mu^2]^2}-(2\pi)^3
\delta^3  (p_t-q_t)
	\int \frac{\rd^3 k}{(2\pi)^3}\frac{8\pi\kappa}{(k^2+\mu^2)^2}, \nn \\
\tilde{V}_2&=&-\frac{16\pi}{3}
	\frac{\alpha_{s
eff}^{2}Q_{0}^{2}}{[(p_t-q_t)^2+\mu^2][(p_t-q_t)^2+Q_{0}^{2}]}.
\label{2n}
\vspace{2mm}
\end{eqnarray}

There are two parameters $\kappa$ and $\alpha_{s eff}$ in the
kernel. However, there should some relation between them since when we 
solve the B-S equation (\ref{2j}) numerically the binding energy should 
satisfy the following relation
\begin{equation}
m_{\Lambda_Q}=m_Q+m_D+E_0,
\label{2o}
\vspace{2mm}
\end{equation}
where we have omitted corrections from $O(1/m_Q)$ since we are working 
in the heavy quark limit. From the B-S equation solutions in meson
case it has been found that the values $m_b=5.02$GeV and $m_c=1.58$GeV give
predictions which are in good agreement with experiments
\cite{dai}. Hence in the baryon case we expect
\begin{equation}
m_D+E_0=0.62GeV.
\label{2p}
\vspace{2mm}
\end{equation}

On the other hand the dimesion of $\kappa$ is three and that of
$\kappa'$ is two. This extra dimesion in $\kappa$ should be caused by
nonperturbative diagrams which include the frozen form factor $F(Q^2)$ 
at low momentum region. Since $\Lambda_{QCD}$ is the only parameter
which is related to confinement we expect that 
\begin{equation}
\kappa \sim \Lambda_{QCD}\kappa'.
\label{2q}
\vspace{2mm}
\end{equation}

Therefore in our numerical calculations we let $\kappa$ vary in the
region between 0.02GeV$^3$ to 0.1GeV$^3$. The diquark mass $m_D$ is
chosen to vary from 650 MeV to 800 MeV. In order to satisfy the relation
(\ref{2p}) we obtain the parameters for different values of $m_D$. The 
results for $m_D=700$MeV are shown in Table 1.

\begin{center}
{\bf Table  1.  Values  of  $\kappa$  and  $\alpha_{s eff}$ for  
$m_D=700$MeV }
\end{center}
\begin{center}
\begin{tabular}{|c|c|c|c|c|c|} 
\hline
$\kappa$(GeV$^3$) &0.02 &0.04 &0.06 &0.08 &0.1
\\ \hline
$\alpha_{s eff}$  &0.67  &0.70 &0.72 &0.74 &0.75
\\ \hline
\end{tabular}
\end{center}
\vspace{2mm}

When $m_D=650$MeV we find that $\alpha_{s eff}=0.60$ for
$\kappa=0.02$GeV$^3$ and $\alpha_{s eff}=0.71$ for
$\kappa=0.1$GeV$^3$. When $m_D=800$MeV, $\alpha_{s eff}=0.79$ for
$\kappa=0.02$GeV$^3$ and $\alpha_{s eff}=0.84$ for
$\kappa=0.1$GeV$^3$. Having these two parameters we obtain the
numerical solution for the B-S wave function directly. 

\vspace{0.2in}
{\large\bf III. The Isgur-Wise function for $\Lamb \rightarrow \Lamc$}
\vspace{0.2in}

In this section we will apply the B-S equation for heavy baryons to 
obtain the numerical result for the Isgur-Wise function for $\Lamb
\rightarrow \Lamc$. The transition diagram is plotted in Fig.2.
In the limit $m_{b,c} \rightarrow \infty$ there is 
only one form factor, the Isgur-Wise function to describe the weak
transition from $\Lamb$ to $\Lamc$,
\begin{equation}
<\Lamc (v')|\bar{c}\gamma_\mu b|\Lamb (v)>=\xi
(\omega)\bar{u}_{\Lamc}(v')\gamma_\mu u_{\Lamb}(v),
\label{3a}
\vspace{2mm}
\end{equation}
where $\omega=v\cdot v'$ is the velocity transfer and $u_{\Lamb}$ and 
$u_{\Lamc}$ are the Dirac spinors of $\Lamb$ and $\Lamc$ respectively.

On the other hand the transition matrix element of $\Lamb \rightarrow
\Lamc$ is related to the B-S wave functions of $\Lamb$ and $\Lamc$ by
the following equation
\begin{equation}
<\Lamc (v')|\bar{c}\gamma_\mu b|\Lamb (v)>=\int
\frac{\rd^4p}{(2\pi)^4} \bar{\chi}_{P'}(p')\chi_{P}(p)S_{D}^{-1}(p_2),
\label{3b}
\vspace{2mm}
\end{equation}
where $P$ ($P'$) is the momentum of $\Lamb$ ($\Lamc$).
$\bar{\chi}_{P'}(p')$ is the wave function of the final state $\Lamc
(v')$ which satisfies the constraint 
\begin{equation}
\bar{\chi}_{P'}(p)\rlap/v'=\bar{\chi}_{P'}(p).
\label{3c}
\vspace{2mm}
\end{equation}
The scalar part of the final state B-S wave function obeys the same
B-S equation as (\ref{2h}). Then from eqs. (\ref{3a}) and (\ref{3b}) 
one gets immediately
\begin{equation}
\xi(\omega)=\int
\frac{\rd^4p}{(2\pi)^4}\phi_{P'}(p')\phi_{P}(p)S_{D}^{-1}(p_2).
\label{3d}
\vspace{2mm}
\end{equation}

Since in the weak transition the diquark acts as a spectator its
mementum in the initial and final baryons should
be the same, $p_2=p'_2$. Then we can show that
\begin{equation}
p'=p+m_D(v'-v),
\label{3e}
\vspace{2mm}
\end{equation}
where again we omitted the $O(1/m_Q)$ corrections. From eq. (\ref{3e}) 
we can get the relations between $p'_l$, $p'_t$ and $p_l$, $p_t$
straightforwardly
\begin{eqnarray}
p'_l &=& p_l\omega -p_t \sqrt{\omega^2-1}cos\theta, \nn \\
p^{'2}_{t} &=& p_{t}^{2}+p_{t}^{2}(\omega^2-1)cos^2\theta
+p_{l}^2(\omega^2-1) -2p_l p_t \omega \sqrt{\omega^2-1} cos\theta,
\label{3f}
\vspace{2mm}
\end{eqnarray}
where $\theta$ is the angle between $p_t$ and $v'_t$.

Substituting the relation between $\phi_P(p)$ and
$\tilde{\phi}_P(p_t)$ (\ref{2k}) into eq. (\ref{3d}) and after integrating
the $p_l$ component by selecting the proper contour we have
\begin{eqnarray}
\xi(\omega) &=& \int \frac{\rd^3
p_t}{(2\pi)^3}\frac{1}{2W_p(E_0+m_D-W_p)(E_0+m_D-\omega W_p
-p_t\sqrt{\omega^2-1}cos\theta)} \nn \\
& & \cdot \int \frac{\rd^3 r_t}{(2\pi)^3}[\tilde{V}_1(p'_t-r_t)-2(\omega W_p
+p_t\sqrt{\omega^2-1}cos\theta)\tilde{V}_2(p'_t-r_t)]\mid _{k_l=-W_p}
\tilde{\phi}_{P'}(r_t) \nn \\ 
& & \cdot \int \frac{\rd^3 l_t}{(2\pi)^3}[\tilde{V}_1(p_t-l_t)-2\omega W_p
\tilde{V}_2(p_t-l_t)]\tilde{\phi}_{P}(l_t), 
\label{3g}
\vspace{2mm}
\end{eqnarray}
where the relation (\ref{3f}) has been substituted into the above equation. 

In eq. (\ref{3g}) all the three-dimensional integral can be
simplified to one-dimensional integral. Furthermore,
the normalization constant of the B-S wave function should
be chosen such that the Isgur-Wise function is equal to one at zero 
momentum transfer. In Fig. 3 we plot numerical result for the
Isgur-Wise function with the
parameter $m_D=700$ MeV. For other values of $m_D$ in the range of
650MeV to 800MeV the shape of the Isgur-Wise function does not change
a lot. 

The slope of the Isgur-Wise function at $\omega=1$ which is of
particular interests is defined as 
\begin{equation}
\frac{\rd\xi(\omega)}{\rd \omega}\mid _{\omega=1}=-\rho^2.
\label{3h}
\vspace{2mm}
\end{equation}

In Table 2 we list the values of $\rho^2$ for different $m_D$ and
$\kappa$. The values of $\rho^2$ outside (inside) the braket
correspond to $\kappa=0.02$GeV$^3$ (0.1GeV$^3$) respectively.

\begin{center}
{\bf Table 2.  Values  of  $\rho^2$}
\end{center}
\begin{center}
\begin{tabular}{|c|c|c|c|} 
\hline
$m_D$(GeV) &0.65 &0.7 &0.8
\\ \hline
$\rho^2$  &1.4 (2.4)  &1.4 (2.4) &1.6 (2.4) 
\\ \hline
\end{tabular}
\end{center}
\vspace{2mm}

It can be seen from Table 2 that the slope is insensitive to the values 
of $m_D$.

The Isgur-Wise function has also been calculated in other models. In
Ref. \cite{guo} Guo and Kroll use the Drell-Yan type overlap integrals 
for the model hadronic wave functions of $\Lamb$ and $\Lamc$ to obtain 
the following form of the Isgur-Wise function
\begin{equation}
\xi(\omega)= \left(\frac{2}{\omega+1}\right)exp\left(-2\varepsilon^2
b^2\frac{\omega-1}{\omega+1}\right) \frac{K_6(2\varepsilon
b/\sqrt{\omega+1})}{K_6(\sqrt{2}\varepsilon b)},
\label{3i}
\vspace{2mm}
\end{equation}
where $\varepsilon$ is the light scalar diquark mass and b is related
to the average transverse momentum of the constituents in the heavy
baryon. $K_l$ is defined as 
$$K_l (x)=\int_{-x}^{\infty}\rd z e^{-z^2}(z+x)^l.$$
The Isgur-Wise function obtained by Jenkins, Manohar and Wise
\cite{soliton} from the soliton model has the following form
\begin{equation}
\xi(\omega)= 0.99 exp[-1.3(\omega-1)].
\label{3j}
\vspace{2mm}
\end{equation}

The MIT bag model calculation by Sadzikowski and Zalewski
\cite{mit}) gives the following result
\begin{equation}
\xi(\omega)= \left(\frac{2}{\omega+1}\right)^{3.5+1.2/\omega}.
\label{3k}
\vspace{2mm}
\end{equation}

The values of $\rho^2$ from the above models are listed in the
following
\begin{eqnarray}
\rho^2=\left \{
            \begin{array}{ll}
            2.9 (3.7)\cite{guo}& \mbox{ for $b=1.18 (1.77)GeV^{-1}$,}\\
            1.3\cite{soliton}, & \mbox{} \\
	    2.4\cite{mit}. & \mbox{}
            \end{array}
                                                 \right.     
\label{3l}
\vspace{2mm}
\end{eqnarray}

We can see that different models may give results for the
Isgur-Wise function for $\Lamb \rightarrow \Lamc$ with some
difference. 
$\rho^2 \geq 1.3$ in all these model calculations. The $\rho$
parameter of the Isgur-Wise
function for $B\rightarrow D$ has also been calculated. In the B-S equation 
approach \cite{dai} $\rho^2$ varies from 1.0 to 1.1.
Other model calculations for $\rho$ in
$B\rightarrow D$ transition gives the following values: $\rho^2=1.19\pm
0.25 $\cite{neubert1}; $\rho=1.13\pm 0.11 $\cite{neubert2}; and $\rho=1.20\pm
0.17 $\cite{rosner}. Because of the large uncertainties we can not
draw a definite conclusion by comparing the slopes of the Isgur-Wise
fuction at the zero recoil point in the meson and baryon cases. However,
it seems that the Isgur-Wise function drops faster in the baryon case 
than in the meson case. At least in the B-S approach the $\rho$
parameter is bigger in the baryon case than in the meson case.

The decay width for the semileptonic transition $\Lamb \rightarrow
\Lamc l \bar{\nu}$ can also be predicted by using the Isgur-Wise function
obtained. The differential decay width has the following form
\cite{guo} \cite{korner}
\begin{equation}
\frac{\rd \Gamma}{\rd
\omega}=\frac{2}{3}
m_{\Lamc}^4 m_{\Lamb}A
\xi^2(\omega)\sqrt{\omega^2 -1}[3\omega(\eta +\eta^{-1})-2-4\omega^2],
\label{3m}
\vspace{2mm}
\end{equation}
where $\eta=m_{\Lamb}/m_{\Lamc}$ and
$A=\frac{G_{F}^2}{(2\pi)^3}|V_{cb}|^2B(\Lamc \rightarrow ab)$. 
$|V_{cb}|$ is the Kobayashi-Maskawa matrix element. $B(\Lamc
\rightarrow ab)$ is the branching ratio for the decay $\Lamc
\rightarrow a({\frac{1}{2}}^+)+ b(0^-)$ through which $\Lamc$ is
detected since the structure for such decay is already well known. 
The plot for $A^{-1}\frac{\rd \Gamma}{\rd
\omega}$ is shown in Fig. 4 for $m_D=700$MeV. Again for other values
of $m_D$ the results change only a little.

After integrating $\omega$ in eq. (\ref{3m}) we have the total decay
width for $\Lamb \rightarrow\Lamc l \bar{\nu}$. For $m_D=700$MeV
\begin{eqnarray}
\Gamma(\Lamb\rightarrow\Lamc l \bar{\nu})=\left \{
            \begin{array}{ll}
            4.2B(\Lamc \rightarrow ab)\times10^{10}s^{-1} 
& \mbox{ when $\kappa=0.02GeV^{3}$,}\\
            5.7B(\Lamc \rightarrow ab)\times10^{10}s^{-1} 
& \mbox{ when $\kappa=0.1GeV^{3}$.}
            \end{array}
                                                 \right.     
\label{3n}
\vspace{2mm}
\end{eqnarray}

When $m_D=650$MeV, $\Gamma(\Lamb\rightarrow\Lamc l \bar{\nu})$ is 
$4.3 (5.9)B(\Lamc \rightarrow ab)\times10^{10}s^{-1}$ for 
$\kappa=0.02(0.1)$ and when
$m_D=800$MeV, $\Gamma(\Lamb\rightarrow\Lamc l \bar{\nu})$ is 
$4.1 (5.0)B(\Lamc \rightarrow ab) \times10^{10}s^{-1}$ for 
$\kappa=0.02(0.1)$. 

\vspace{0.2in}
{\large\bf IV. Summary and discussions}
\vspace{0.2in}

In this paper we established the B-S equation for the
heavy baryon which is considered as composed of a heavy quark and a
light scalar diquark. By assuming the kernel containing scalar
confinement and one-gluon-exchange terms we solve the B-S equation
numerically. Furthermore, we applied the obtained results to calculate
the Isgur-Wise function  for $\Lamb \rightarrow \Lamc$.
It is found that in the B-S approach, the Isgur-Wise function for
$\Lamb \rightarrow \Lamc$ decreeses faster than for $B\rightarrow
D$. Comparison with other model calculations are also presented. The
slope parameter $\rho^2$  at the zero recoil point for $\Lamb
\rightarrow \Lamc$ is larger than 1.3
in all the present model calculations. 
In the limit
$m_{b,c}\rightarrow \infty$ we also calculated the differential and
total decay widths for $\Lamb\rightarrow\Lamc l \bar{\nu}$.

Although the B-S equation is formally the exact equation to describe
the bound state, there is much difficulty in applying it to the real
physical state. The most difficult point is the that we can not solve
out the form of the kernel. Hence we have to use some phenomenological
kernel which is reasonable because of the success of the potential
model. This leads to some uncertainties. In our approach,
the parameters $\kappa$ and $\alpha_{s eff}$ in the kernel are not
exactly determined. Therefore, we let them to vary in some resonable 
range. The experimental data concerning $\Lambda_Q$ in the future can
help to fix the parameters in our model. 

We have worked in the heavy quark limit. The physical
predictions such as $\Gamma(\Lamb\rightarrow\Lamc l \bar{\nu})$
will be subjective to $1/m_Q$ corrections. The study on the $1/m_Q$
corrections will appear in our later work.

\vspace{1cm}

\noindent {\bf Acknowledgment}:
\vspace{2mm}

One of us (X.-H. G) would like to thank the Japan
Society for the Promotion of Science (JSPS) for the support. He 
is also indebted to Dr. H.-Y. Jin for helpful
discussions.

\newpage

\vspace{0.2in}

{\large \bf Figure Captions} \\
\vspace{0.2in}

Fig.1 The diquark-gluon-diquark vertex.\\
\vspace{0.2cm}

Fig.2 The weak transition diagram for $\Lamb \rightarrow \Lamc$.\\
\vspace{0.2cm}

Fig.3 	The Isgur-Wise function for $\Lamb \rightarrow
\Lamc$($m_D=700$MeV). For the solid line $\kappa=0.02$GeV$^3$ and for the
dotted line $\kappa=0.1$GeV$^3$. \\
\vspace{0.2cm}

Fig.4 	The differential decay width for $\Lamb \rightarrow
\Lamc l \bar{\nu}$ ($m_D=700$MeV). For the solid line
$\kappa=0.02$GeV$^3$  and for the
dotted line $\kappa=0.1$GeV$^3$. \\
\vspace{0.2in}

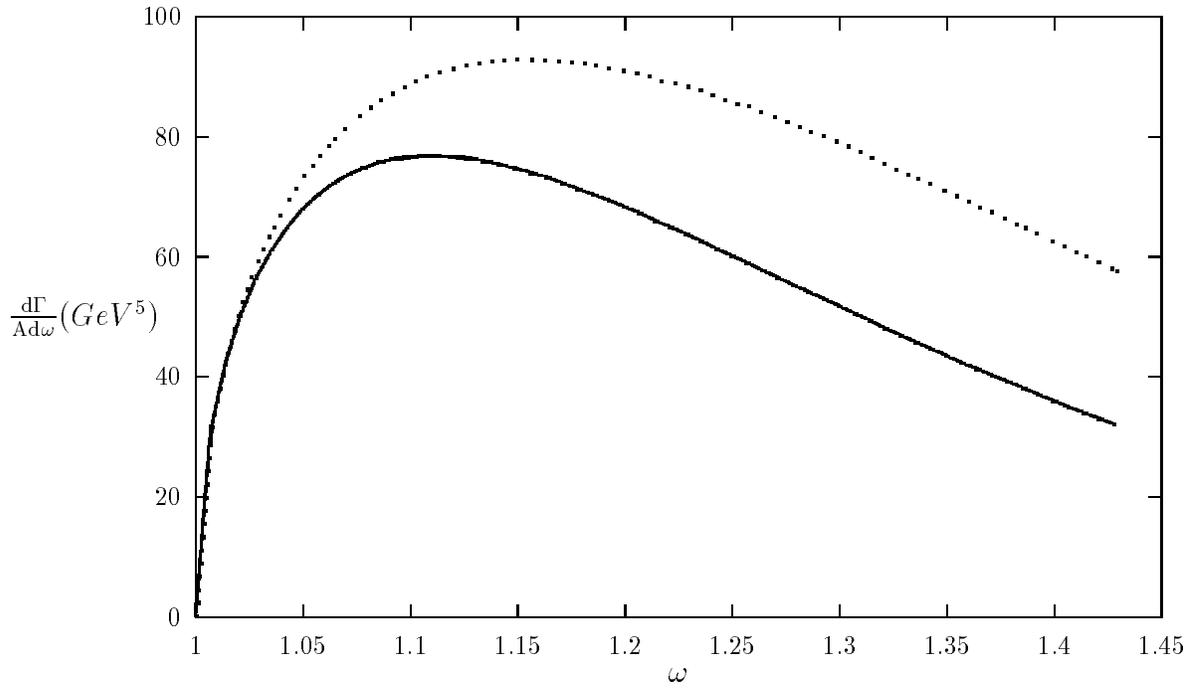
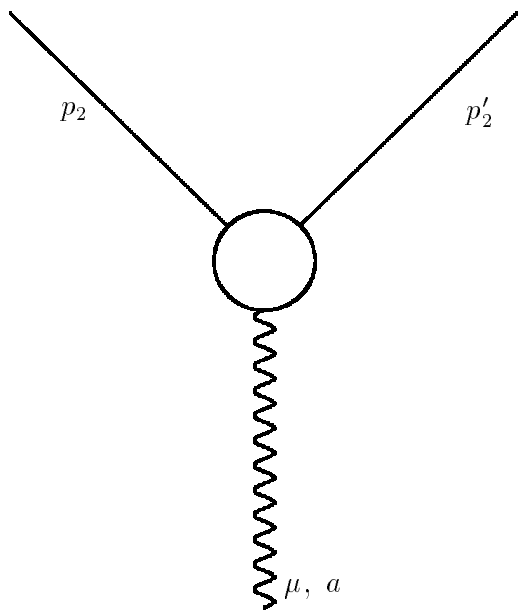
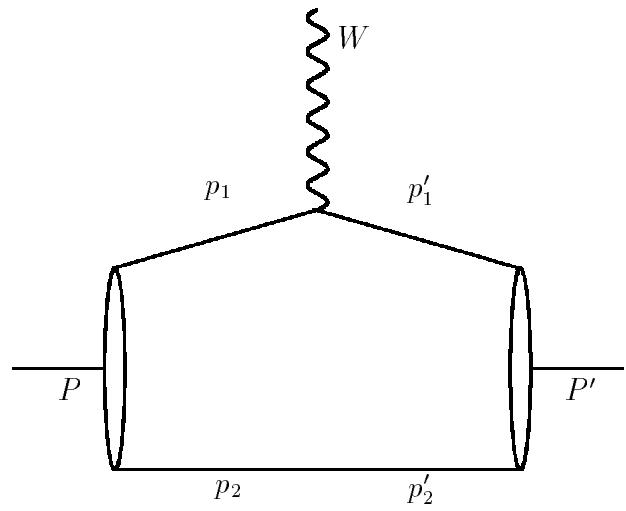
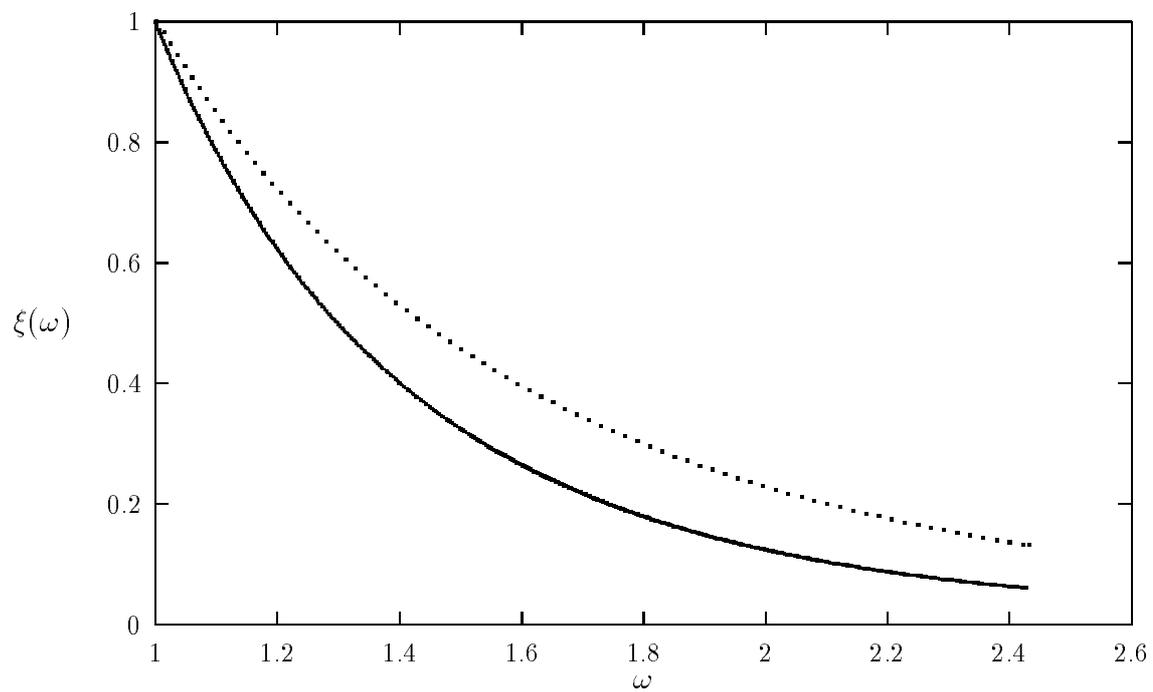
\end{document}